\documentclass[aps,showpacs,superscriptaddress,pra]{revtex4}

\usepackage{amsmath,amssymb}
\usepackage{graphicx}
\usepackage{dcolumn}
\usepackage{bm}
\begin{document}

\title
{Impact of Electric Fields on Highly Excited Rovibrational States of Polar Dimers}

\author{Rosario Gonz\'alez-F\'erez}, 
\affiliation{Instituto 'Carlos I' de F\'{\i}sica Te\'orica y Computacional
and Departamento de F\'{\i}sica At\'omica Molecular y Nuclear,
Universidad de Granada, E-18071 Granada, Spain}
\author{Peter Schmelcher}
\affiliation{Theoretische Chemie, Physikalisch--Chemisches Institut,
Im Neuenheimer Feld 229, D-69120 Heidelberg, Germany}
\affiliation{Physikalisches Institut, Universit\"at Heidelberg,
Philosophenweg 12, D-69120 Heidelberg, Germany}
\email{rogonzal@ugr.es, Peter.Schmelcher@pci.uni-heidelberg.de}

\date{\today}

\begin{abstract}
We study the effect of a strong static homogeneous electric field on the
highly excited rovibrational levels of the LiCs dimer in its electronic
ground state. Our full rovibrational investigation of the system includes
the interaction with the field due to the permanent electric dipole moment
and the polarizability of the molecule. We explore the evolution of the
states next to the dissociation threshold as the field strength is
increased. The rotational and vibrational dynamics are influenced
by the field; 
effects such as orientation, angular motion hybridization and squeezing of
the vibrational motion are demonstrated and analyzed. The field also
induces avoided crossings causing a strong mixing of the electrically
dressed rovibrational states. Importantly, we show how some of these
highly excited levels can be shifted to the continuum as the field
strength is increased, and reversely how two atoms in the continuum can be
brought into a bound state by lowering the electric field strength.
\end{abstract}


\maketitle

\section{\label{sec:intro}Introduction}

The recent availability of cold to ultracold polar dimers in the vibrational
and  rotational ground state of their singlet electronic ground potential
\cite{ni08,deiglmayr:133004}, represents a breakthrough towards the control of
all molecular degrees of freedom, i.e., the center of mass,
electronic, rotational and vibrational motions, and towards the ultimate goal
of obtaining a polar condensate.      
The experimental achievements have been accompanied by significant theoretical
efforts to understand the intriguing physical phenomena expected 
for ultracold polar quantum gases due to their anisotropic and long-range
dipole-dipole interaction. In particular it has been analyzed how external fields can control and
manipulate the scattering properties 
\cite{gorshkov:073201,ticknor:133202,tscherbul:194311,avdeenkov:022707},
and the chemical reactions dynamics \cite{krems_pccp}, or how to use them as tools for
quantum computational devices \cite{demille02,yelin06,zoller_qc}. 
Different approaches to achieve cold and ultracold molecules have been explored 
\cite{pellegrini:053201,cote06,gonzalez:023402,gonzalez07_2,kotochigova07}. 
For further work and references we refer the reader to the comprehensive reviews \cite{speciss2004_2,dulieu:jpb_39}.

The most widespread techniques to produce ultracold polar molecules are the 
photoassociation of two ultracold atoms \cite{stwalley99,jones:483}, and the
tuning of the atomic interactions via magnetically induced Feshbach resonances 
\cite{Koehle:rmp06}. 
Alternative pathways explore the ability to manipulate 
the interaction between atoms by inducing optical Feshbach resonances. Based on
the same principle as the magnetically induced Feshbach resonances, they appear
when two colliding ultracold atoms are coupled to a bound state of the
corresponding molecular system by using a radiation field. 
Initially, there were several theoretical proposals to obtain these resonances with the
help of radio-frequency, static electric, and electromagnetic fields   
\cite{PhysRevLett.77.2913,PhysRevA.56.1486,PhysRevLett.81.4596,kokoouline:3046}.
In addition, it has been demonstrated that a combination of a magnetic and
static electric field can induce Feshbach resonances in a binary mixture of
Li and Cs atoms \cite{krems06,li:032709},
and that a suitable combination of these two fields can tune the relevant
interaction parameters, such as the width and open-channel scattering length
in these resonances \cite{marcelis:153201}. 
The existence of this optically induced resonances has been experimentally
proved for different atomic species by tuning the laser frequency near a
photoassociation resonance  
\cite{PhysRevLett.85.4462,PhysRevLett.93.123001,thalhammer:033403,enomoto:203201}.

Within the above experimental techniques, the molecules are usually in a highly
excited vibrational level close to the dissociation threshold of an electronic  
state. These vibrational states are exposed to external fields and  most of their
overall probability is located at the outermost hump of their  probability 
density. 
In the present work, we investigate the few last most weakly bound states of the
$\textrm{X}^1\Sigma^+$ electronic ground state of a polar molecule in  a
strong static electric field.   
We perform a full rovibrational investigation of the field-dressed nuclear
dynamics including the interaction of the field with the molecular electric dipole
moment and polarizability. The LiCs dimer is a prototype system and will be used here. This choice is
based on the experimental interest in this system and the availability of
its molecular polarizabilities \cite{deiglmayr:064309}.
It completes our previous investigations on the effects of an electric
field on this system, where we have analyzed the rovibrational spectrum, the
radiative decay properties, and the formation of these ultracold dimers 
via single-photon photoassociation from the continuum into its electronic ground state
\cite{gonzalez07_2,gonzalez:023402, gonzalez06,mayle06,gonzalez08}.
Specifically, we analyze the binding energies and the expectation values,
$\langle\cos\theta\rangle$, $\langle \mathbf{J}^2\rangle$ and $\langle R\rangle$,
of states lying in the spectral region with binding energies 
smaller than $0.28$ cm$^{-1}$ and vanishing azimuthal quantum number in the
very strong field regime. 
We demonstrate that both the rotational and vibrational dynamics are
significantly affected by the field. Indeed, the vibrational motion is
squeezed or stretched to minimize the energy, depending on the rotational degree of excitation
and the field strength. At such strong fields the nuclear spectrum exhibits
several avoided crossings 
between energetically adjacent states, which lead to a strongly distorted
rovibrational dynamics. The latter might be directly observable when imaged by photodissociation experiments.
Beyond this, (magnetically induced) avoided crossings in Cs$_2$ have been used
to  construct a molecular St\"uckelberg interferometer \cite{mark:113201}.  
In addition, we show that by tuning the electric field strength a dissociation
channel is opened, i.e., a weakly bound molecular state with low-field seeking
character is shifted to the atomic continuum by increasing the field
strength. Of course, the reverse process is also possible, and two free atoms
can be brought into a molecular bound state by lowering the field strength.   
This might be of interest to control the collisional dynamics of the atomic/molecular cold gas
by using either very strong static (micro-) electric fields or strong
quasistatic, i.e., time-dependent fields. 

\section{\label{sec:theo} The Rovibrational Hamiltonian} 
We consider a heteronuclear diatomic molecule in its $^1\Sigma^+$ electronic
ground state exposed to a homogeneous and static electric field.  
Our study is restricted to a non-relativistic treatment and 
addresses exclusively a spin singlet electronic ground state, and therefore relativistic
corrections can be neglected. We assume that for the considered regime of field
strengths perturbation theory holds for the description of the interaction
of the field with the electronic structure, whereas a nonperturbative
treatment is indispensable for the corresponding nuclear dynamics.  
In addition, we take into account the interaction of the field with the molecule via its dipole
moment and polarizabibility, thereby neglecting higher order contributions due to (higher)
hyperpolarizabilities. Thus, in the framework of the
Born-Oppenheimer approximation the rovibrational Hamiltonian reads  
\begin{equation}
\label{eq:rotvib_hamiltonian}
H= T_R
+\frac{\hbar^2\mathbf{J}^2(\theta,\phi)}{2\mu R^2} + 
V(R)-FD(R)\cos\theta
-\frac{F^2}{2}
\left[\alpha_\bot(R)\sin^2\theta+\alpha_\parallel(R)\cos^2\theta\right], 
\end{equation}
where $R$ and $\theta, \phi$ are the internuclear distance and the Euler
angles, respectively, and we use the molecule fixed 
frame with the coordinate origin at the center of mass of the nuclei.  
$T_R$ is the vibrational kinetic energy, $\hbar\mathbf{J}(\theta,\phi)$ is the
orbital angular momentum, $\mu$ is the reduced mass of the nuclei, and 
$V (R)$ is the field-free electronic potential energy curve (PEC). 
The electric field is taken oriented along the $z$-axis of the laboratory
frame  with strength $F$. 
The last three terms provide the interaction between the electric field and
the molecule via its permanent electronic dipole moment function 
(EDMF) $D(R)$, and its polarizability, with 
$\alpha_\parallel(R)$ and $\alpha_\bot(R)$ being the polarizability components
parallel and perpendicular to the molecular axis, respectively.

In the presence of the electric field, the dissociation threshold changes and
it is given by the quadratic Stark shift of the free atoms, i.e.,
$E_{DT}(F)=-0.5*F^2(\alpha_{1}+\alpha_{2})$, 
with $\alpha_{i}$ $i=1,2$ being the  polarizabilities of the free atoms. 
In the presence of the electric field, only the azimuthal symmetry of the
molecular wavefunction holds and therefore the magnetic quantum number 
$M $ is retained. In this work we focus on levels with vanishing magnetic quantum number $M=0$.
For reasons of addressability, we will label the electrically-dressed
states by means of their field-free vibrational and rotational quantum numbers
$(\nu,J)$.

Let us briefly investigate under which conditions the contribution of the
molecular polarizabilities can be neglected 
in the Hamiltonian (\ref{eq:rotvib_hamiltonian}). 
For simplicity reasons, and without loss of generality, we use the
effective rotor approach \cite{gonzalez04}, assuming that the rotational and
vibrational energy scales differ significantly and can
therefore be separated adiabatically, and  that the field influence on the
vibrational motion is very small and can consequently be treated by
perturbation theory. 
Then, in the framework of this approximation the rovibrational Hamiltonian
(\ref{eq:rotvib_hamiltonian}) is reduced to 
\begin{equation}
\label{hamil_era}
 H_\nu^{ERA}=
B_\nu\mathbf{J}^2 -F\langle D\rangle_\nu^{0}\cos\theta 
-\frac{F^2}{2}\left[\langle\alpha_\bot\rangle_\nu^{0}+
\langle \Delta\alpha\rangle_\nu^{0}\cos^2\theta\right] +E_{\nu00}^{0},
\end{equation}
where $B_\nu =\frac{\hbar^2}{2\mu}\langle R^{-2}\rangle_\nu^{0}$ is the
field-free rotational constant of the state with quantum numbers $\nu$, $J=0$
and $M=0$, $\psi_{\nu00}^{0}(R)$ and $E_{\nu00}^{(0)}$ are the
vibrational wave function and energy, respectively, and we encounter the
expectation values  
$\langle R^{-2}\rangle_\nu^{0}=\langle\psi_{\nu 0 0}^{0}|R^{-2}|\psi_{\nu 0 0}^{0}\rangle$,
$\langle D \rangle_\nu^{0} = \langle\psi_{\nu 0 0}^{0}|D(R)|\psi_{\nu 0 0}^{0}\rangle$, 
$\langle\alpha_\bot\rangle_\nu^{0}=
\langle\psi_{\nu 0 0}^{0}|\alpha_\bot(R)|\psi_{\nu 0 0}^{0}\rangle$, and 
$\langle \Delta\alpha\rangle_\nu^{0}=
\langle \psi_{\nu 0 0}^{0}|\alpha_\parallel(R)-\alpha_\bot(R)|\psi_{\nu 0 0}^{0}\rangle$. 
Within this approach, at a certain field strength $F$ the interaction due to
the polarizability can be neglected in the effective rotor Hamiltonian
(\ref{hamil_era}) if  
$\left|
\frac{2\langle D\rangle_\nu^{0}}{F\langle \alpha_\bot \rangle_\nu^{0}}
\right|>>1$ 
and  
$\left|
\frac{2\langle D\rangle_\nu^{0}}{F\langle \Delta\alpha\rangle_\nu^{0}}
\right|>>1$. 

To analyze the very weak field regime, we rescale the effective rotor 
Hamiltonian (\ref{hamil_era}) with $B_\nu$, and assume that the ratios
$\frac{F}{B_\nu}\langle D \rangle_\nu^{0}$, 
$\frac{F^2}{2B_\nu}\langle \alpha_\bot  \rangle_\nu^{0}$, and 
$\frac{F^2}{2B_\nu}\langle \Delta\alpha \rangle_\nu^{0}$ are smaller
than the rescaled field-free rotational kinetic energy $J(J+1)$.
Then, for a certain state with quantum numbers $\nu$, $J$ and $M$, 
time-independent perturbation theory provides the following second order
correction to the field-free energy  
\begin{equation}
  \label{eq:pertur_the}
  E^{(2)}_{\nu,J,M}=\left[
A_{JM}\frac{\left(\langle D \rangle_\nu^{0}\right)^2}{B_\nu} 
-\frac{1}{2}\langle \alpha_\bot  \rangle_\nu^{0}
-\frac{1}{2}\langle \Delta\alpha \rangle_\nu^{0}C_{JM}\right]F^2, 
\end{equation}
where the angular coefficients \cite{meyenn} are given by 
 \begin{eqnarray}
   \label{eq:p_t_dipole}
  A_{JM}&=&\frac{J(J+1)-3M^2}{2J(J+1)(2J+1)(2J+3)} 
\quad  \rm{for} 
  \quad J>0 \nonumber \\
   A_{00}&=&-\frac{1}{6} 
 \end{eqnarray}
and 
\begin{equation}
  \label{eq:p_t_pola}
  C_{JM}=\frac{(J+1)^2-M^2}{(2J+1)(2J+3)}+ \frac{J^2-M^2}{(2J+1)(2J-1)}.
\end{equation}
 
This second order correction to the rotational energy depends on the
molecular system and the symmetry of the considered state through the
expectation values $\langle D \rangle_\nu^{0}$, $\langle \alpha_\bot
\rangle_\nu^{0}$ and $\langle \Delta\alpha \rangle_\nu^{0}$ .
In the perturbative regime, the polarizability terms can be neglected
if   
$\left|
\frac{2(\langle D\rangle_\nu^{0})^2A_{JM}}{B_\nu\langle \alpha_\bot \rangle_\nu^{0}}
\right|>>1$ 
and  
$\left|
\frac{2(\langle D\rangle_\nu^{0})^2A_{JM}}
{B_\nu\langle \Delta\alpha\rangle_\nu^{0}C_{JM}} 
\right|>>1$. 
We have $C_{JM} > |A_{JM}|$, and the coefficient $|A_{JM}|$ becomes increasingly
smaller than $C_{JM}$ for increasing values of $J$; for example 
for $J=15$ and $M=0$ $A_{15,0}=4.888\times 10^{-4}$ and $C_{15,0}=0.5005$,
and for $J=M=15$ $A_{15,15}=-8.859\times 10^{-4}$ and 
$C_{15,15}=3.030\times 10^{-2}$. As a consequence we encounter the situation that
for high rotational excitations of certain molecular systems the interaction
due to the molecular polarizability could be the dominant one.   
We emphasize that the above considerations are valid only for weak fields and within the effective
rotor approach.

\section{\label{sec:result} Results} 

\begin{figure}
\includegraphics[scale=0.8]{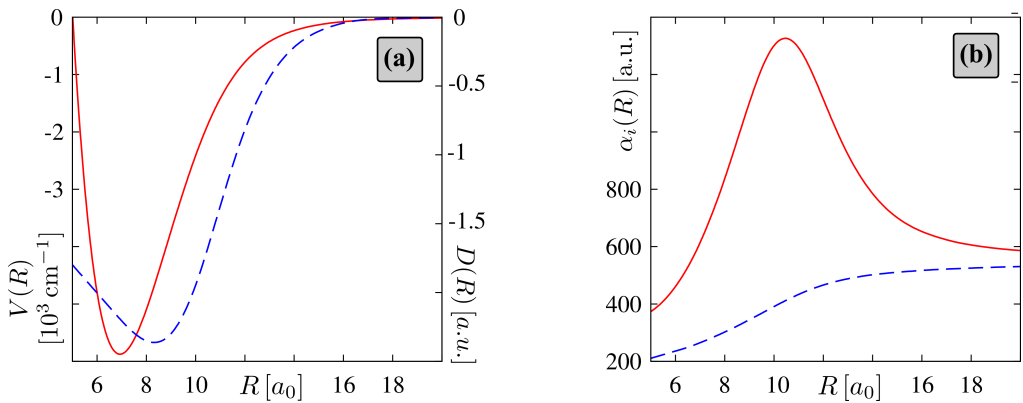}
\caption{\label{fig:pec}
(a): Electronic potential curve (solid) and electric dipole moment functions
(dashed),  and (b): parallel $\alpha_\parallel(R)$ (solid) and 
perpendicular $\alpha_\bot(R)$ (dashed) components of
the polarizability of the electronic ground state of the LiCs molecule.}
\end{figure}

In the present work, we have performed a full rovibrational study of the
influence of an external static electric field on the highly excited rovibrational states of the LiCs molecule. 
The PEC, EMDF and polarizability components of the $^1\Sigma^+$ electronic
ground state of LiCs are plotted as a function of the internuclear distance in
Figures \ref{fig:pec}(a), and (b), respectively.
For the PEC, we use the experimental data of ref.\cite{staanum06}, which includes for the long-range behaviour the van der
Waals terms, $-\sum_{n=6,8,10}C_n/R^{n}$, and an exchange energy term,
$-AR^\gamma e^{-\beta R}$, see ref.\cite{staanum06} for the values of these parameters. 
The EDMF and polarizabilities are taken from semi-empirical calculations 
performed by the group of Dulieu \cite{deiglmayr:064309,aymar05}.
The EDMF is negative and its minimum is shifted by $1.4\,a_0$ with respect
to the equilibrium internuclear distance $R_e=6.94\,a_0$ of the PEC. 
For the electronic ground state of the polar alkali dimers the long-range
behaviour of the EDMF is given by $D_7/R^{7}$ \cite{Byers1970}, this function 
has been fitted to the theoretical data for $R\gtrsim 18.15\,a_0$ with 
$D_7=-5\times 10^{-6}$ a.u. 
Regarding the polarizability, both components smoothly change as 
$R$ is enhanced and $\alpha_\bot(R)\ge\alpha_\parallel(R)$ for any $R$
value.
They satisfy that 
$\lim_{R\to\infty}\alpha_\bot(R)=\lim_{R\to\infty}\alpha_\parallel(R)=
\alpha_{Li}+\alpha_{Cs}$,  
with the polarizabilities of the Li and Cs atoms 
$\alpha_{Li}=164.2$ a.u. and $\alpha_{Cs}=401$ a.u., respectively, 
\cite{miffre:011603,PhysRevLett.91.153001}.
Thus,  for $R\gtrsim 26$ a$_0$ the theoretical data were extrapolated by means
of exponentially decreasing functions to match the constant value 
$\alpha_{Li}+\alpha_{Cs}$.  
For computational reasons, $\alpha_\bot(R)$ and $\alpha_\parallel(R)$ are
extrapolated for $R<5$ and $4$ a$_0$, respectively.  
Since this study is focused on highly excited levels lying close to the
dissociation threshold, we are aware of the fact that our results strongly
depend on the assumptions made for the long-range behaviour of $D(R)$, 
$\alpha_\bot(R)$ and $\alpha_\parallel(R)$, and on the extrapolations
performed at short-range for $\alpha_\bot(R)$ and $\alpha_\parallel(R)$.
However, let us remark that the overall behaviour and physical
phenomena presented here remain unaltered as these parameters are altered.

For the lowest rotational excitations within each vibrational band of LiCs,
we have investigated and compared the field interactions with 
the dipole moment and polarizability presented in the previous section. 
Within perturbation theory, the interaction due to the molecular
polarizability becomes comparable to the one due to the dipole moment only for the last
two vibrational bands. Assuming that the effective rotor conditions are
satisfied, the interaction with the polarizability can be neglected for those
levels with $\nu\le 47$ and $48\le\nu\le 52$ if the field strength is smaller
than  $10^{-3}$ and $2\times 10^{-4}$ a.u., respectively. Whereas, for the
vibrational bands $\nu=53$ and $54$ both interactions possess the same order of
magnitude for the much weaker fields $F\approx 6\times 10^{-5}$ and $8\times 10^{-6}$ a.u.,
respectively. 
Furthermore, the absolute values of the quadratic Stark shifts of the atomic 
energies are larger than the binding energies of the last bound state for
$F\gtrsim 10^{-5}$ a.u., which also justifies that the interaction with the
polarizability has to be included in the present study.  

Here, we consider the highest rotational excitations ($M=0$) for the last four
vibrational bands, $51\le\nu\le54$, of LiCs with binding energies smaller than
$0.28$ cm$^{-1}$. We focus on the strong field regime 
$F = 10^{-6}-3.4\times 10^{-4}$ a.u., i.e., $F = 5.14- 1747.6$ kV/cm, which includes
the experimentally accessible range of strong static fields and possibly quasistatic fields.
We remark that such strong fields are considered to induce the below-described peculiar behaviour
of these states. Most of the overall probability is located at the outermost
hump of these states, i.e., in regions where the EDMF possesses
small values and the polarizabilities are close to $\alpha_{Li}+\alpha_{Cs}$. Thus, 
strong fields are needed in order to observe a significant field-effect on
these levels. 
At these field strengths the corresponding rovibrational dynamics cannot be
described by means of the effective or (due to avoided crossings) even the adiabatic rotor approximations
\cite{gonzalez04,gonzalez05}, and, of course not by perturbation theory (\ref{eq:pertur_the}).  
Hence, the two-dimensional Schr\"odinger equation associated to the 
nuclear Hamiltonian (\ref{eq:rotvib_hamiltonian}) has to be solved numerically.
We do this by employing a hybrid computational method which combines discrete
and basis-set techniques applied to the radial and angular 
coordinates, respectively  \cite{gonzalez:023402,gonzalez04}.

\begin{figure}
\includegraphics[scale=0.7]{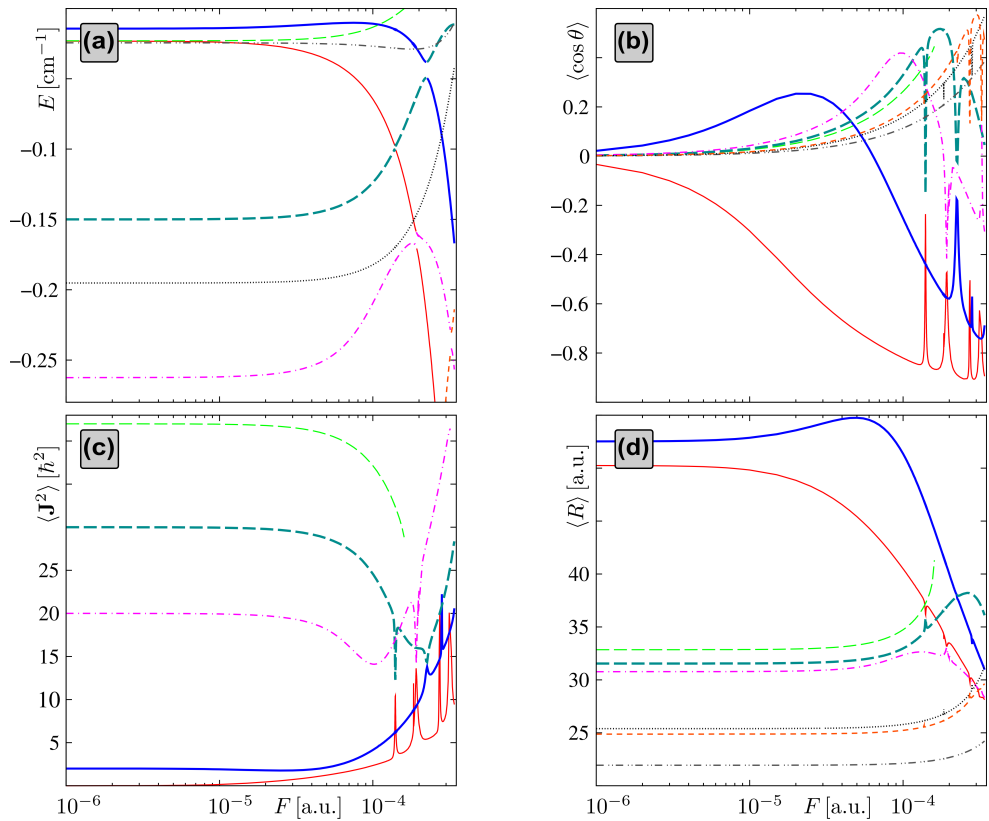}
\caption{\label{fig:energy}
(a) Energy shifts with respect to the dissociation threshold 
$\varepsilon_{\nu J}$, and expectation values (b) $\langle\cos\theta\rangle$,
(c) $\langle\mathbf{J}^2\rangle$ and (d) $\langle R\rangle$, as a function of
the field strength for the states  
with field-free vibrational and rotational quantum numbers $(54,0)$ (solid),
$(54,1)$ (solid thick), $(53,4)$ (dotted-dashed), $(53,5)$ (long dashed thick),
$(53,6)$ 
(long dashed), $(52,10)$  (dotted), $(52,9)$ (short dashed),
 and $(51,15)$ (double dotted-dashed). Note that in the panel (c) only the results
for the levels  $(54,0)$, $(54,1)$, $(53,4)$, $(53,5)$ and $(53,6)$ are
included.}
\end{figure}

Since in the presence of the field the dissociation threshold is 
$E_{DT}(F)=-0.5*F^2(\alpha_{Li}+\alpha_{Cs})$, we define the
energetical shift with respect to this dissociation threshold as 
$\varepsilon_{\nu, J}=E_{\nu, J}(F)-E_{DT}(F)$, with $E_{\nu, J}(F)$ being the 
energy of the $(\nu,J)$ state at field strength $F$. 
Figure \ref{fig:energy}(a) shows these Stark shifts $\varepsilon_{\nu, J}$ 
satisfying $ \varepsilon_{\nu, J} \ge -0.28$ cm$^{-1}$ 
in the above-provided range of field strengths.  
This spectral window includes the $(54,0)$, $(54,1)$, $(53,4)$, $(53,5)$, 
$(53,6)$, $(52,10)$ and $(51,15)$ states, and for $F\gtrsim 3 \times 10^{-4}$
a.u. also the $(52,9)$ level.  
Since all these levels possess the same symmetry ($M=0$) for $F\neq 0$, and since
the field strength is the only parameter at hand to vary the rovibrational
energies, the von Neumann-Wigner noncrossing rule  \cite{wigner} holds and we
encounter only avoided crossings of energetically adjacent states  but no
exact crossings in the field-dressed spectrum.   
In the vicinity of the avoided crossing the nuclear dynamics is dominated by a strong
interaction and mixing of the involved rovibrational states. For reasons of simplicity
(according to the Landau-Zener theory) we assume that the avoided crossings are 
traversed diabatically as $F$ is increased. Thus, a
certain state has the same character before and after the avoided crossing, 
e.g. taking for example the $(54,0)$ level we observe that it keeps its
high-field seeking trend, see Figure \ref{fig:energy}(a).

Before studying these avoided crossings in more detail, let us analyze
the general behaviour of the binding energies. 
For all $\varepsilon_{\nu J}$ we observe a very weak dependence on $F$ 
for $F \lesssim 3\times 10^{-5}$ a.u. The larger the field-free rotational
quantum 
number of a state, the stronger is the field strength needed to encounter a
deviation  of $\varepsilon_{\nu, J}$  from its field-free value.  
With a further enhancement of the strength, $\varepsilon_{\nu, J}$ increases   
(decreases) for the  high (low)-field seekers. 
The strong field dynamics is dominated by pendular states, whose binding
energies increase as $F$ is augmented. Their main feature is their
orientation along the field axis: They represent coherent superpositions of
field-free rotational levels \cite{meyenn}.     
In our spectral region, this regime is only reached for the $(54,0)$,
$(54,1)$ and $(53,4)$ states. 
Indeed, $\varepsilon_{54,\, 0}$ monotonically decreases as $F$ increase, 
whereas $\varepsilon_{54,\, 1}$  and $\varepsilon_{53,\, 4}$ initially
increase and  reach broad maxima, decreasing thereafter. 
In contrast, the binding energies of the $(52,9)$, $(52,10)$, $(53,6)$ and
$(53,5)$ decrease as $F$ is increased.
Due to its large field-free angular momentum, the $(51,15)$ level is the least
affected by the field. Initially, $\varepsilon_{51,\, 15}$ is reduced as $F$
is enhanced, passes through a broad minimum and increases thereafter. 
The contribution of the molecular polarizability causes that this
state is a high-field seeker in the weak field regime, whereas if only the
interaction with the dipole moment is included it has a low-field seeking
character.  

Regarding the avoided crossings, some of them are very narrow and can not be
identified as such on the scale used in Figure \ref{fig:energy}(a), 
e.g. those among the pairs of levels $(54,1)-(53,6)$, $(54,1)-(51,15)$,
$(54,0)-(51,15)$,  and $(54,0)-(52,10)$. 
In contrast, other avoided crossings are very broad, and they are
characterized by a strong coupling between the involved molecular states. 
The avoided crossing between the $(54,1)$ and $(53,5)$ levels
takes places at strong fields, and the minimal energetical gap is     
$\Delta E=|\varepsilon_{54,\,1}-\varepsilon_{53,\,5}|=1.13\times 10^{-2}$
cm$^{-1}$ for $F\approx 2.23 \times 10^{-4}$ a.u. 
The $(54,0)$ state is involved in an avoided crossing with the $(53,5)$ level
characterized by  
$\Delta E=|\varepsilon_{54,\,0}-\varepsilon_{53,\,5}|=2.82\times 10^{-3}$ 
cm$^{-1}$ 
for $F\approx 1.399\times 10^{-4}$ a.u., and another one with the $(53,4)$
state with 
$\Delta E=|\varepsilon_{54,\,0}-\varepsilon_{53,\,4}|=1.08\times 10^{-2}$ 
cm$^{-1}$ for $F\approx 1.918 \times  10^{-4}$ a.u.
The $(53,4)$ level experiences an avoided crossing with the $(52,9)$ state,
with  minimal energetical separation  
$\Delta E=|\varepsilon_{53,\,4}-\varepsilon_{52,\,9}|=1.12\times 10^{-2}$ 
cm$^{-1}$ for $F\approx 3.25 \times  10^{-4}$ a.u.
For other alkali dimers weaker electric field strengths might suffice
to exhibit similar avoided crossings.
These avoided crossings are wide enough to be experimentally observed in a
similar way as it has been done for the weakly bound spectrum of 
Cs$_2$ dimer in a magnetic field \cite{mark:042514}. 
In this system coupling strengths, $\frac{\Delta E}{2}$, larger than 
$1.67\times 10^{-6}$ cm$^{-1}$ were experimentally estimated; i.e., even 
the energetical separation of the $(54,1)$ and $(53,6)$ avoided crossing,  
$\Delta E=|\varepsilon_{54,\,1}-\varepsilon_{53,\,6}|=9.6\times 10^{-5}$ 
cm$^{-1}$ for $F\approx 1.1 \times 10^{-4}$ a.u., could be measured.
Moreover, as it has been done for the Cs$_2$ molecule
\cite{mark:042514,mark:113201} in a magnetic field, a suitable 
electric-field ramp could be used to transfer population from high to low
rotational excitations in a controlled way, by either
diabatically jumping or adiabatically following these electrically induced
avoided crossings.  

An interesting physical phenomenon is observed in the evolution of the
$(53,6)$ level in the spectrum.  
$\varepsilon_{53,\, 6}$ increases as $F$ increases, and
after passing the avoided crossing with the $(54,1)$, 
$\varepsilon_{53,\, 6}$ becomes positive for $F\gtrsim 1.6\times 10^{-4}$a.u.  
The Stark increase of the $(53,6)$ energy surpasses the
reduction of the dissociation threshold, 
and this level is shifted to the continuum. 
Hence, if the LiCs is initially in the $(53,6)$ level, it will dissociate
as the field strength is adiabatically tuned and enhanced above 
$F\gtrsim 1.6\times 10^{-4}$ a.u. 
Therefore a channel for molecular dissociation is opened  as the electric field is
modified.  
Of course, the inverse process is also possible, and the continuum state
formed by two free atoms can be brought into a bound state by lowering the
electric field strength.  
Indeed, it has been proved that a static electric field could be used to
manipulate the interaction between two atoms such that a virtual state
could be transformed in a new bound state, i.e., the molecular system supports
a new bound level \cite{PhysRevLett.81.4596}.

To illustrate the appearance of this phenomenon for a low-lying rotational 
excitation, we have performed a similar study for a designed molecule.
We have taken the theoretical PEC of the $^1\Sigma^+$ electronic ground state 
of LiCs computed by the group of Allouche \cite{allouche} with the van der
Waals long-range potential, $C_6/R^6$, but modifying the LiCs $C_6$ coefficient
to $C_6=2225$ a.u. 
The $(54,1)$ level is shifted towards the dissociation threshold, having a
field-free energy $ E_{54\,1}\approx -5.9\times 10^{-5}$ cm$^{-1}$.  
As electric dipole moment function and polarizabilities we have used the
corresponding functions of the LiCs molecule described above.  
The last most weakly bound states of this toy system have been studied in the
presence of a static electric field, but for the sake of simplicity we
discuss here only the results for the $(54,1)$ level.
As $F$ is enhanced $\varepsilon_{54,\, 1}$ increases, and becomes positive
for $F\gtrsim 5\times 10^{-5}$ a.u.; note that this field strength is much weaker
than the above used ones. For a bound level, a further enhancement of the field would change its
character, and its binding energy would increase as $F$ is augmented. 
We have observed the same phenomenon for the $(54,1)$ state, which
becomes bound again for $F\gtrsim 1.7\times 10^{-4}$ a.u., and 
$\varepsilon_{54,\,1}$ decreases thereafter.
The level has been captured by the nuclear potential demonstrating that the
reverse process is possible.
Starting with two free atoms with the correct internal symmetry, by
adiabatically tuning the field the dimer is formed in a highly excited level. 

Due to negative sign of the EDMF, the main feature of the pendular regime (focusing again on
LiCs) is the antiparallel orientation of the states along  the field axis. The
orientation can be estimated by the expectation value 
$\langle\cos\theta\rangle$: The closer $|\langle\cos\theta\rangle|$ is to one,
the stronger is the orientation of the state along the field.
Figure \ref{fig:energy}(b) illustrates the evolution of 
$\langle\cos\theta\rangle$ as the field strength is changing.
The initial behaviour of $\langle\cos\theta\rangle$ for weak fields depends on the character
of the corresponding level.   
For the $(54,0)$ state $\langle\cos\theta\rangle$ monotonically decreases
as $F$ is increased, it achieves the largest orientation with   
$\langle\cos\theta\rangle\le-0.7$ for $F\gtrsim 5 \times 10^{-5}$ a.u., except
in the proximity of avoided crossings.
For the $(54,1)$, $(53,5)$ and $(53,4)$ levels, $\langle\cos\theta\rangle$
reach a broad maximum decreasing thereafter. The orientation of the $(54,1)$
and $(53,4)$ states becomes antiparallel for stronger fields.
Not considering the proximity of an avoided crossing region, the $(54,1)$
state shows a significant orientation with 
$\langle\cos\theta\rangle\le-0.4$ for $F\gtrsim 1.31 \times 10^{-4}$ a.u. 
The remainder of states keep a pinwheeling character, and
$\langle\cos\theta\rangle$ increases as $F$ is augmented.  
Since we have used the notation that the avoided crossings are traversed diabatically,
a certain state $\langle\cos\theta\rangle$ reestablishes its increasing or
decreasing trend once the avoided crossing has been passed.
The smooth behaviour of $\langle\cos\theta\rangle$ is significantly distorted
by the presence of these spectral features, where due to the strong mixing and 
interaction among the two involved states $\langle\cos\theta\rangle$ exhibits
sharp and pronounced maxima and minima. 
For example, the avoided crossing among the $(54,0)$ and $(53,5)$ levels, 
is characterized by the values $\langle\cos\theta\rangle_{54,\,0}=-0.235$ and 
$\langle\cos\theta\rangle_{53,\,5}=-0.152$, for $F=1.399\times 10^{-4}$ a.u.,
compared to the results $\langle\cos\theta\rangle_{54,\,0}=-0.847$ and 
$\langle\cos\theta\rangle_{53,\,5}=0.436$ obtained 
for $F=1.3\times 10^{-4}$ a.u.
Note that for the $(54,0)$ level $\langle\cos\theta\rangle$ shows an
additional maximum for $F\gtrsim 3\times 10^{-4}$, i.e., this level suffers
another avoided crossing which is not observed in Figure \ref{fig:energy}(a),
because $\varepsilon_{54,0}<-0.28$ cm$^{-1}$ for $F\ge 2.54\times 10^ {-4}$ a.u. 

The expectation value $\langle\mathbf{J}^2\rangle$ of the states $(54,0)$,
$(54,1)$, $(53,4)$, $(53,5)$ and $(53,6)$, is presented as a function
of the electric field in Figure \ref{fig:energy}(c). To provide a reasonable scale,
the results for the $(52,10)$, $(52,9)$ and $(51,15)$ levels have not been
included.  
This quantity provides a measure for the mixture of field-free states with
different rotational quantum numbers $J$ but the same value for $M$, i.e., it
describes the hybridization of the field-free rotational motion.
Analogous to the binding energy, $\langle\mathbf{J}^2\rangle$ shows for weak fields
a plateau-like behaviour: The hybridization of the angular motion is
very small and the dynamics is dominated by the field-free rotational quantum
number of the corresponding state.
For stronger fields, these states possess a rich rotational dynamics, with
significant contributions of different partial waves, and 
$\langle\mathbf{J}^2\rangle$ decreases (increases) for the low-(high)-field
seekers as $F$ is enhanced. 
In the strong field regime, $\langle\mathbf{J}^2\rangle$ shows a broad
minimum for the $(54,1)$, $(53,4)$ and $(53,5)$ states, increasing thereafter.
The pendular limit is characterized by the augment of 
$\langle\mathbf{J}^2\rangle$ due to the contribution of higher field-free
rotational states. This regime is only achieved by the $(54,0)$, $(54,1)$ and
$(53,4)$ levels.   
In contrast, the mixing with lower rotational excitations is dominant 
for the $(53,5)$ and $(53,6)$ states, and
$\langle\mathbf{J}^2\rangle\le J(J+1)$, with $J$ being the corresponding
field-free rotational quantum number; similar results are obtained for the
$(52,9)$, $(52,10)$ and $(51,15)$ states not included in Figure 
\ref{fig:energy}(c).
The presence of the avoided crossings significantly distorts the smooth
behaviour of $\langle\mathbf{J}^2\rangle$. 
The $\langle\mathbf{J}^2\rangle$ of the level in an avoided crossing with the
lowest (highest) field-free $J$ exhibits 
a pronounced and narrow maximum (minimum) on these irregular regions. 
At the smallest energetical gap, we encounter similar values of
$\langle\mathbf{J}^2\rangle$ for both states. For example, for 
$F=1.399\times 10^{-4}$ a.u. we obtain 
$\langle\mathbf{J}^2\rangle=10.38\,\hbar^2$ and $12.40\,\hbar^2$ for the
$(54,0)$ and $(53,5)$ levels, respectively, compared to the values
$\langle\mathbf{J}^2\rangle_{54,0}=3.04\,\hbar^2$ and
$\langle\mathbf{J}^2\rangle_{53,5}=20.75\,\hbar^2$ for $F=1.3\times 10^{-4}$ a.u.

The expectation value of the radial coordinate $\langle R\rangle$ is presented
for these states and range of field strengths in Figure \ref{fig:energy}(d). 
Only if the vibrational motion is affected by the field
$\langle R\rangle$ should differ from its field-free value.
Analogously to $\varepsilon_{JM}$ and $\langle\mathbf{J}^2\rangle$, 
$\langle R\rangle$ represents approximately a constant for weak fields, and 
strong fields are needed to observe significant deviations from its  
field-free value.
Indeed, the larger is the rotational quantum number of a state for $F=0$,  the
least affected by the field is its $\langle R\rangle$.
For the $(54,0)$ level, $\langle R\rangle$ monotonically decreases from 
$50.24\, a_0$ to $28.13\,a_0$ as $F$ is enhanced from $0$ to 
$3.4 \times 10^{-4}$ a.u.  
For the $(54,1)$, $(53,5)$ and $(53,4)$ states, $\langle R\rangle$ increases
as $F$ is augmented, reaches a broad maximum and decreases thereafter. The
$(54,1)$ level is significantly affected with a reduction from
$\langle R\rangle=52.52\, a_0$ to $30.96\, a_0$ for $F=0$ and $
3.4\times 10^{-4}$ a.u., respectively. For the $(53,4)$ state
this effect is much smaller, and $\langle R\rangle$ is modified 
from the field-free result $30.78\, a_0$ to $28.32\,a_0$. 
For $(53,5)$ we observe that for $F=3.4 \times 10^{-4}$ a.u.
$\langle R\rangle$ is by $4.57\,a_0$ larger than its value for $F=0$.
$\langle R\rangle$ increases as $F$ is enhanced
for the remaining states, their total rise being smaller than $5\, a_0$ for the
analyzed levels $\nu=51$ and $52$. 
As the $(53,6)$ state is shifted to the continuum, the slope of 
$\langle R\rangle$ becomes very steep, 
and $\langle R\rangle$ is enhanced from $\langle R\rangle=32.85\, a_0$ up to
$41.30\,a_0$ for $F=0$ and $1.6\times 10^{-4}$ a.u., respectively. 
The field effect on the vibrational motion can be  explained as follows: 
The probability density of those levels with a antiparallel (parallel)
orientation is mostly located in the  $\pi/2\le\theta\le \pi$   
($0\le\theta\le \pi/2$) region, where the dipole moment interaction is
attractive (repulsive). As a consequence, the wavefunctions are squeezed
(stretched) compared to their field-free counterparts to reduce the energy.   
Again, in the vicinity of the avoided crossing $\langle R\rangle$ exhibits
very similar values for the two involved states. 
For example, we have found that $\langle R\rangle=33.02$ and $33.13\, a_0$ for
the states $(54,0)$ and $(53,5)$ and $F=1.399\times 10^{-4}$ a.u., respectively.

 \begin{figure}
 \includegraphics[scale=0.8]{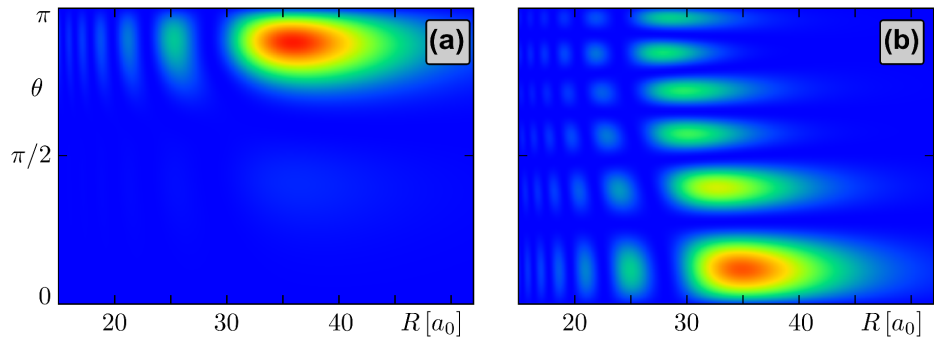}
 \caption{\label{fig:wf_1}
 Probability densities (a) of the state $(54,0)$ and (b) of the state
  $(53,5)$ for $F=1.3\times 10^{-4}$ a.u., i.e., close to the avoided crossing
   but still  without mixing of the rovibrational field-dressed states.}
 \end{figure}
To gain a deeper insight into the coupling of the vibrational
and rotational motions induced by the electric field, we have analyzed the
corresponding wavefunctions of two states involved in an  avoided crossing.  
As an example, we discuss here the $(54,0)$-$(53,5)$ avoided crossing. 
For comparison, let us first analyze their wavefunctions for 
$F=1.3\times 10^{-4}$ a.u., i.e., 'below' the avoided crossing
where the mixing is not yet appreciable. The contour plots of the probability densities, 
$|\psi(R,\theta)|^2\sin \theta$, in the $(R,\theta)$ plane are presented in
Figures \ref{fig:wf_1}(a) and (b) for the $(54,0)$ and $(53,5)$ states, 
respectively.  
Since most of the overall probability of these weakly bound levels is located
in the outer most hump, the radial coordinate has been restricted in these
plots to the interval $15 \,a_0\le  R\le 52\,a_0$.   
Indeed, more than $89\%$ of the $(54,0)$ and $(53,5)$  probability densities
are located for $R>30\,a_0$, and $25\, a_0$, respectively.
Due to the pronounced antiparallel orientation of the $(54,0)$ level,
$\langle\cos\theta\rangle =-0.847$, the corresponding probability 
density shows a pendular-like structure, it is located in the region
$3\pi/4\le\theta\le\pi$ and the maximal value is obtained at $\theta=2.77$ and
$R=35.87\,a_0$.  
The typical oscillator-like behaviour with $6$ maxima reminiscent from its
field free angular momentum $J=5$ is observed in the $(53,5)$ probability
density, see Figure \ref{fig:wf_1}(b).
Since this state has still a pinwheeling character, the corresponding
probability density is distributed over the complete interval 
$0\le\theta\le \pi$, however, due to the parallel orientation of this state,
$\langle\cos\theta\rangle=0.437$, the probability density is larger in the region
$\theta\le\frac{\pi}{2}$. 
Moreover, the influence of the field on the vibrational motion provokes an
inclination of the internuclear axis of this level, i.e., the corresponding 
wavefunction is stretched and squeezed in the regions $\theta<\frac{\pi}{2}$ and
$\theta>\frac{\pi}{2}$, respectively. The squeezing effect also appears for the $(54,0)$ state 
which possesses a strong antiparallel orientation (see Figure
\ref{fig:wf_1}(a)).

 \begin{figure}
  \includegraphics[scale=0.8]{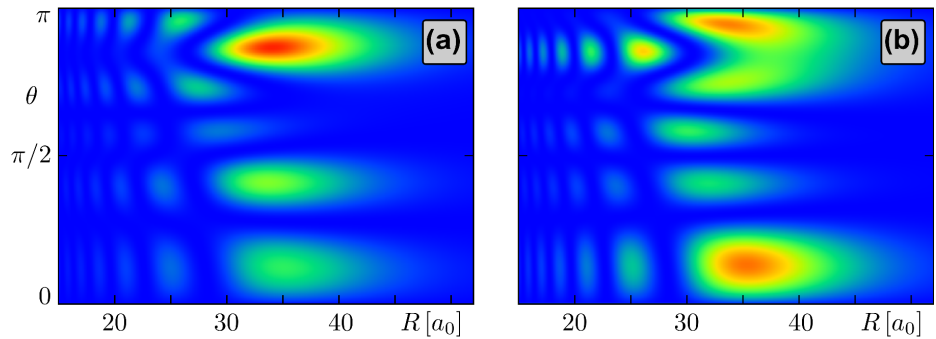}
  \caption{\label{fig:wf_2}
  Probability densities (a) of the state $(54,0)$ and (b) of the state
    $(53,5)$  at the field strength $F=1.399\times 10^{-4}$ a.u.}
  \end{figure}
As the electric field is enhanced approaching the region of the avoided crossing, a
strong interaction between the involved states takes places and 
the rovibrational dynamics is affected drastically. The contour plots of the $(54,0)$ and $(53,5)$ states for $F=1.399 \times 10^{-4}$ a.u.,
which corresponds with the minimal energetical gap between them, are shown in
Figures \ref{fig:wf_2}(a) and (b), respectively.  
Although, at this field strength their orientation and hybridization of the 
angular motion are very similar, there exist significant differences with respect to their wavefunctions.  
The above-described regular structures typical for an oscillator and
pendular-like 
distributions are lost. Even more, for both levels it is not possible to identify an orientation of the molecule,
and the most pronounced maxima are not
necessarily located at the outermost turning points.
The $(54,0)$ probability density is distributed in the interval
$0\le\theta\le \pi$, the largest maximum is at $\theta=2.701$ and
$R=34.02\,a_0$, and it is 
accompanied by several less pronounced maxima at smaller $\theta$ values. 
The $(53,5)$ probability density exhibits three maxima with similar
probability density, the first one at $\theta=0.44$ and $R=35.24\,a_0$, 
the second one at $\theta=2.95$ and $R=33.73\,a_0$, and the third one, at
$\theta=2.70$ and $R=25.97\,a_0$, which is shifted towards smaller internuclear
separations from the outermost turning point.  
Both configurations exhibit significantly distorted patterns, and they show a strong mixing
between the radial and angular degrees of freedom. In general, at the avoided
crossings the nuclear dynamics of the field-dressed states are characterized
by an asymmetric and strongly distorted behavior, exhibiting pronounced
localization phenomena.

We have also analyzed these weakly bound levels taking into account only the
interaction of the field with the permanent electric dipole, and not
considering the contribution of the molecular polarizability. The results look
qualitatively similar but  show a quantitatively different behaviour as a
function of the electric field, and the effect of the polarizability becomes
important for $F \gtrsim 10^{-4}$ a.u. 
The polarizability terms cause the mixing of states with field-free
rotational quantum numbers $J$ and $J\pm2$. 
If the polarizability is included, the dissociation energies are
smaller, i.e., for a certain $F$ value the modulus of the displacement of
the dissociation threshold is larger than the modulus of the energetical shift
due to polarizability of a certain level, and the avoided crossings are also
broader. Without the contribution of the polarizability term the
$(53,6)$ level is not shifted to the continuum as the field is increased,
and the inverse phenomenon appears, i.e., the $(54,2)$ level becomes a
bound state for $F\gtrsim 1.8\, \cdot 10^{-4}$ a.u.  

The validity of the adiabatic and effective rotor approaches has been
previously demonstrated for vibrational low-lying levels of the LiCs
dimer \cite{gonzalez06}. However, this does not hold true for the 
part of the spectrum considered here. Since both approximations do not include the full coupling
between the vibrational and rotational motion, the 
presence of the avoided crossings in the spectrum is not reproduced.
In addition, significant errors are found for the binding energies and the
expectation value $\langle R\rangle$ of the rotational excitations even in
the absence of the field, e.g. the $(53,6)$ and $(51,15)$ levels are not
bound  within these approaches.  
The above-discussed field-effects on the vibrational motion
can not be explained using an effective rotor description  \cite{gonzalez04}.
However, the adiabatic results qualitatively reproduce the orientation and
hybridization of the angular motion as well as the stretching and
squeezing of the vibrational motion. Numerically significant deviations
are encountered in the avoided crossing regions.

\section{\label{sec:conclu} Conclusions and outlook} 
We have investigated the influence of a strong static and homogeneous electric
field on the highly excited rovibrational states of the electronic ground state X$^1\Sigma^+$
of the alkali dimer LiCs by solving the fully coupled rovibrational
Schr\"odinger equation. The interaction of the field with
the electric dipole moment function as well as with the molecular
polarizability has been taken into account. We focus here on 
the last rotational excitations with vanishing azimuthal
symmetry within the last four vibrational bands,  $51\le\nu\le 54$.
Due to their large extension, strong fields are needed in order to observe a
significant field influence. 

The richness and variety of the resulting field-dressed rotational dynamics has been
illustrated by analyzing the energetical Stark shifts, as well as,
the orientation, the hybridization of the angular motion and the
vibrational stretching and squeezing effects. 
Whether we encounter a squeezing or stretching of the vibrational motion depends on the angular
configuration: The molecule tries to minimize its energy leading to stretching
for a parallel configuration and squeezing for an antiparallel one.
In the strong field regime, the electrically-dressed spectrum is characterized
by the presence of pronounced avoided crossings between energetically adjacent levels.  
These irregular features lead to a strong field-induced 
mixing and interaction between the states, and they cause strongly distorted
and asymmetric features of the corresponding probability densities.
We stress the importance of identifying these irregular features: Their
presence affects the radiative decay properties of the dimer, such as lifetime
and transition probability for spontaneous decay, and they   
might significantly alter the chemical reaction dynamics. Even more, one of
their possible applications is their use to transfer population between the
involved states.   

We have demonstrated that if the last most weakly bound state is a
low-field seeker it is possible to shift it to the atomic continuum by tuning
the electric field, i.e., the molecular system dissociates into free atoms.    
The reverse process is also possible, i.e., a continuum state, formed by two free
atoms with the correct field-free rotational symmetry, can be transferred to a
weakly bound molecular state by changing the field strength. 
These results suggest that by properly increasing or decreasing the value of
the electric field, one could study in a control way the opening of a
dissociation or an association channel.

Although our study is restricted to a LiCs dimer and to the spectral region
close to the dissociation threshold of its electronic ground state, we
stress that the above-observed physical phenomena are expected to occur in many other polar
molecules.

\acknowledgments
Financial support by the Spanish projects FIS2008-02380 (MEC) and
FQM--0207, FQM--481 and P06-FQM-01735 (Junta de Andaluc\'{\i}a) is
gratefully appreciated. This work was partially supported by the National
Science Foundation through a grant for the Institute for Theoretical Atomic,
Molecular and Optical Physics at Harvard University and Smithsonian
Astrophysical Observatory. Financial support by the Heidelberg Graduate School
of Fundamental Physics in the framework of a travel grant for R.G.F. is gratefully
acknowledged. We thank Michael Mayle for his help with respect to technical
aspects of this work.

\bibliography{molecules}
\end{document}